\documentclass[aps,prb,twocolumn,showpacs]{revtex4}
\usepackage{bm}
\usepackage{amsmath}
\usepackage{graphicx}

\newcommand{\Imp}{{\text{Im}}\,}

\newcommand{\beq}{\begin{equation}}
\newcommand{\eeq}{\end{equation}}
\newcommand{\beqar}{\begin{eqnarray*}}
\newcommand{\eeqar}{\end{eqnarray*}}

\newcommand{\Mintr}{M_{F0}}
\newcommand{\Mtot}{M_{\text{tot}}}
\newcommand{\ETh}{E_{\text{Th}}}
\newcommand{\fBCS}{f_S}
\newcommand{\gBCS}{g_S}
\newcommand{\sgn}{\,\text{sgn}\,}

\newcommand{\ua}{\uparrow}
\newcommand{\da}{\downarrow}

\newcommand{\pd}{\partial}
\newcommand{\rtarr}{\rightarrow}

\newcommand{\gc}{\check{g}}

\newcommand{\ac}{\check{a}}
\newcommand{\sch}{\check{s}}

\newcommand{\eA}{\epsilon_A}

\newcommand{\lan}{\langle}
\newcommand{\ran}{\rangle}
\newcommand{\om}{\omega}
\newcommand{\Om}{\Omega}
\newcommand{\al}{\alpha}

\newcommand{\ga}{\gamma}

\newcommand{\de}{\delta}
\newcommand{\De}{\Delta}

\newcommand{\sig}{\sigma}

\newcommand{\lt}{\left}
\newcommand{\rt}{\right}
\begin{document}

\title{Oscillations of Induced Magnetization in
Superconductor-Ferromagnet Heterostructures}
\author{M. Yu. Kharitonov$^{1}$, A. F. Volkov$^{1,2},$ K. B. Efetov$^{1,3}$}
\affiliation{$^{1}$ Ruhr-Universit\"{a}t Bochum, Germany,\\
$^{2}$Institute of Radioengineering and Electronics of RAS, Moscow,
Russia.\\
$^{3}$L.D. Landau Institute for Theoretical Physics, Moscow, Russia.}
\date{\today}

\begin{abstract}
We study a change in the spin magnetization of a
superconductor-ferromagnet (SF) heterostructure, when temperature is lowered below
the superconducting transition temperature.
It is assumed that the SF interface is smooth on the atomic
scale and the mean free path is not too short. Solving the
Eilenberger equation we show that the  spin magnetic moment induced in the
superconductor
is an oscillating sign-changing function of the product $hd$ of the  exchange field
$h$ and the thickness $d$ of the ferromagnet.
Therefore the total spin magnetic moment of the system in the superconducting state
can be not only smaller (screening)
but also greater (anti-screening)
than that in the normal state, in contrast with the case of
highly disordered (diffusive) systems,
where only screening is possible.
This surprising effect is due to
peculiar periodic properties of localized Andreev states in the system.
It is most pronounced in systems with ideal ballistic transport
(no bulk  disorder in the samples, smooth ideally transparent interface),
however these ideal conditions are not crucial for
the very existence of the effect.
We show that oscillations exist (although suppressed)
even for arbitrary low interface
transparency and in the presence of bulk disorder, provided that
$h \tau \gg 1$ ($\tau$ -- mean free path).
At low interface transparency we solve the problem
for arbitrary strength of disorder
and obtain oscillating magnetization in ballistic regime ($h \tau \gg 1$)
and nonoscillating magnetization in diffusive one ($h \tau \ll 1$)
as limiting cases of one formula.
\end{abstract}
\maketitle
\section{Introduction\label{sec:intro}}
Spin structures of the microscopic states of
{\em s}-wave superconductors
and ferromagnets are opposite to each other \cite{Abrikosov,SST}.
Superconducting pairing interaction leads to formation of electron Cooper
pairs with opposite projections of spins, whereas the exchange field tends
to align electron spins in the same direction. This counteraction on the
microscopic level results in a competition between macroscopic
superconducting and magnetic states. The suppression of the superconducting
order parameter and the transition temperature by the exchange field \cite{SST,Sarma}
and the reduction of the magnetic spin susceptibility in the
superconductors \cite{Yosida,AG} are well-known examples of this competition.

Among suitable experimental systems for studying the interplay of the
superconductivity and spin magnetism are superconductor-ferromagnet (SF)
heterostructures (for reviews see Refs.~\onlinecite{Golubov,Buzdin,BVErmp}).
Above the superconducting transition temperature $T_c$,
the superconductor is in its normal state, and the total magnetic moment $M_{\text{tot}}$
of such system is given by the intrinsic
magnetic moment of
the ferromagnet $M_{F0}$.
Below $T_c$ a magnetic moment $M(h)$ induced
by the presence of superconductivity appears,
and the total spin magnetic moment of SF system in the superconducting state
is $\Mtot=\Mintr+M(h)$.
The induced magnetization may be caused by both the
Meissner currents (orbital effect) and the spin polarization (spin effect).
If the sizes of the ferromagnet and the superconductor
are small compared to the London penetration length,
then the orbital effect is small compared to the spin effect \cite{BVE}.
In this work we assume this situation,
since we want to study the effect
related to the spin polarization.
Therefore $M(h)$ is the induced {\em spin} magnetic moment throughout the paper.

What direction of $M(h)$ relative to $\Mintr$ one would expect?
The above-mentioned competing behavior of
superconducting and magnetic phenomena
suggests that $M(h)$ is opposite ($M(h)<0$) to $\Mintr$ and thus reduces $\Mtot$.
In other words, the induced magnetization
screens the intrinsic magnetization of the ferromagnet.
The idea of  spin screening of the ferromagnet's magnetization by the superconductor in SF systems
was first brought forward in Refs.~\onlinecite{BVE}.
In these publications the cases of a ferromagnetic planar film and spherical grain
were considered and it was shown that indeed $M(h)<0$.

Experiments carried out on various SF structures confirm indirectly
the idea of the screening proposed in Refs.~\onlinecite{BVE}.
In Ref.~\onlinecite{Garifullin} a V-Pd$_{1-x}$Fe$_{x}$ SF bilayered
structure was studied by a magnetic resonance technique and a $50\%$-decrease
in $M_{\text{tot}}$ was discovered as the temperature was lowered
from $T=T_{c}\approx 4{\text K}$ to $T\approx 1.5{\text K}$.
In Ref.~\onlinecite{Stahn} the
neutron reflectometry was performed on multilayered SF structures consisting
of ferromagnetic La$_{2/3}$Ca$_{1/3}$MnO$_{3}$ and superconducting
YBa$_{2}$Cu$_{3}$O$_{7}$ layers.
The obtained reflectometry spectra were discussed
in context of the screening effect predicted in Ref.~\onlinecite{BVE}.

\begin{figure}
\centering
\includegraphics[width=0.49\textwidth, height=0.25\textwidth]{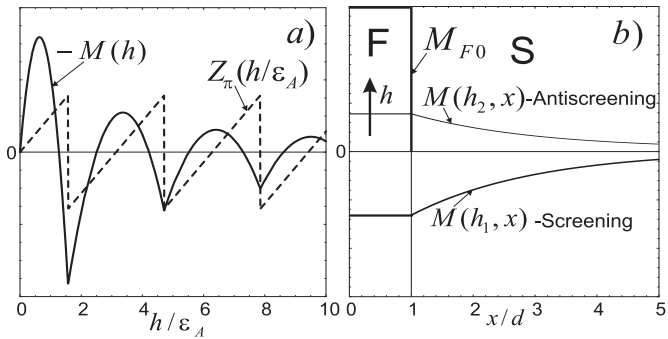}
\caption{(a) Oscillations of induced magnetic moment $M(h)$ of a clean SF system
as a function of $h/\protect\epsilon_A=2hd/v_F$ (solid line shows $-M(h)$).
The graph is plotted for the case of zero temperature $T=0$ and thin F-layer
$d \ll \xi_S$.
At $hd \ll v_F$ we have $M(h)=-\chi_N h d$,
indicating complete screening in the case itinerant ferromagnet.
Dashed graph shows the function $Z_\pi ( h/\epsilon_A )$
(see Eq.~(\ref{eq:MT0})).
(b) The geometry of the system and spatial distribution of the density $M(h,x)$
of the induced magnetization
($h=h_1$ -- screening, $h=h_2$ -- anti-screening).}
\label{fig:M}
\end{figure}

In Refs.~\onlinecite{BVE}
the screening effect was studied under the following assumptions:
i)~``diffusive'' limit
(the mean free path of electrons $l$ is much smaller than both
the size of the ferromagnet $d$ and the superconducting coherence length $\xi_S$);
ii)~the exchange field of the ferromagnet $h$ is small compared to
the Thouless energy $\ETh= D/d^2$
($D$ -- diffusion coefficient in the ferromagnet,
throughout the paper we employ units, in which the Planck constant $\hbar=1$):
\beq
    h \ll \ETh,
\label{eq:hcond}
\eeq
and the ferromagnetic film is thin ($d \ll \xi_S$).
Condition i) allowed to use Usadel equation
and condition ii) to treat the effect of ferromanget's exchange
field $h$ as a pertubation in this equation. For the induced spin magnetic moment
$M(h)$ the following result was obtained:
\beq
    M(h)=-(\chi_N- \chi_S(T))h d,
\label{eq:Msmallh}
\eeq
where $\chi_S(T)$ is the magnetic susceptibility
 of a bulk superconductor,
$\chi_N$ is the magnetic susceptibility of the superconductor in the normal state
($\chi_S(T_c)=\chi_N$) (For exact expression for $\chi_S(T)$, see Eq.~(\ref{eq:chiS})).
The result  Eq.~(\ref{eq:Msmallh}) is quite universal
\footnote{It should be noted that it was assumed
in Refs.~\onlinecite{BVE}
and will be assumed throughout this paper
that no spin-orbit or magnetic
scattering is present in the system; such processes would, of course, affect this result.},
since it is independent
of the strength of potential disorder and
interface transparency.

The following questions arise.
First, how robust is this perturbative result Eq.~(\ref{eq:Msmallh})
to the type of orbital electron dynamics,
in particular, what result would be obtained in
the opposite case of a clean ballistic system.
Second and more interestingly,
how does induced magnetization $M(h)$ behave for sufficiently large exchange field,
when its effect cannot be considered as a perturbation anymore
and how the type of electron dynamics affects this behavior.
The theory presented below shows that the behavior of induced magnetization $M(h)$
in ballistic SF systems in nonperturbative regime can be very remarkable.

First, let us define more precisely what we mean by
``perturbative'' and ``nonperturbative'' regimes for
an SF system in general case.
For a generic SF system
with arbitrary bulk disorder in S and F regions
and arbitrary interface transparency an important energy scale
is  $\epsilon^*=1/\tau^*$, where $\tau^*$ is the characteristic time
spent by electron in the ferromagnet.
(Ferromagnet is assumed to be of finite size $d$ at least in one dimension).
In ballistic system without or with relatively weak bulk disorder
(the mean free path $l \gtrsim d$)
and with not too small interface transparency ($t\sim 1$)
this energy scale is the Andreev
energy: $\epsilon^*=\eA=v_F/d$ ($v_F$ is the Fermi velocity).
In the case of low interface transparency $t\ll 1$ the time $\tau^*$
is enhanced due to the fact that
electron has to hit the interface $\sim 1/t$ times
before it escapes from the ferromagnet,
therefore it stays in the ferromagnet $\sim 1/t$ times longer
compared to the case of good transparency.
Thus, for the case of low interface transparency ($t \ll 1$) one gets
$\epsilon^*=\eA t$ for ballistic system.
In the diffusive system ($l \ll d$) with not too small interface transparency
($t \gg l/d $) $\epsilon^*=\ETh=D/d^2$ is the Thouless energy.

Comparison of $h$ with $\epsilon^*$
determines how strongly the exchange field $h$
affects the spectrum of SF system compared
to the spectrum of the corresponding SN(superconductor-normal metal) system
with $h=0$.If
\beq
    h/\epsilon^* \ll 1,
\label{eq:pert}
\eeq
then the effect of exchange field is small
and consequently  physical quantities, such as the induced magnetization $M(h)$,
can be studied perturbatively. We refer to the regime (\ref{eq:pert})
as perturbative.
On the contrary, when
\beq
    h \gtrsim \epsilon^*,
\label{eq:nonpert}
\eeq
the spectrum of SF system is significantly altered by the presence of exchange field $h$,
and we refer to the regime (\ref{eq:nonpert}) as nonperturbative.

A very interesting property of Andreev spectrum of SF system in nonperturbative regime is
its periodicity as a function of parameter $h/\epsilon^*$
with the period of the order of unity\cite{AndreevSF}.
(The physical reasons beyond this phenomenon and the origin of the conditions
(\ref{eq:pert}),(\ref{eq:nonpert})
are given in Sec.\ref{sec:SFspectrum}.)
This periodicity  is known to reveal itself in the
oscillations of the Josephson critical current $I_c(h)$ in SFS junctions
\cite{JosephTh1,JosephTh2}.
One could expect to find these periodic features
of Andreev spectrum in other macroscopic quantities,
such as the induced magnetization $M(h)$.
In this paper we extended the analysis of Refs.~\onlinecite{BVE}
of induced magnetization in SF structures to nonpertubative case and
found that in ballistic systems  this is indeed the case.

Namely, we considered an SF bilayered system with the ferromagnet of the thickness $d$
and the superconductor of the size much greater than its coherence length $\xi_S$ (Fig.1(b)).
Solving the Eilenberger equation
we find that in a system without or with
relatively weak bulk ($l \gg d$) or surface disorder and with perfect SF interface transparency
 ($t=1$)
the induced spin magnetic moment $M(h)$ is
an {\em oscillating sign-changing function} (Fig. 1(a))  of
parameter $h/\eA=2 h d /v_F$
with the quasiperiod approximately equal to $\pi $.
Therefore, the total magnetic moment
\beq
    M_{\text{tot}}=M_{F0}+M(h)   \label{eq:Mtot}
\eeq can be either smaller ($M(h)<0$) or larger ($M(h)>0$) than
that in the normal state $M_{\text{tot}}^{N}=M_{F0}$ depending on
the value of the product $hd$. The oscillations of $M(h)$ are most
pronounced for the system with ideal transport properties:
ballistic electron motion, perfect interface transparency.
However, these ideal conditions are not crucial for the very
existence of oscillations. To verify this we have considered the
case of low SF interface transparency ($t \ll 1$) and arbitrary
disorder in the ferromagnet, described by the scattering time
$\tau$ The limit of low SF interface transparency ($t \ll 1$) is
useful from a methodological standpoint, since it allows to solve
the Eilenberger equation for the system with arbitrary strength of
disorder. This allows one to study not only the limiting diffusive
and ballistic cases, but also the crossover between them. It
appears that the influence of disorder on the behavior of induced
magnetization $M(h)$ is governed by parameter $h\tau$. In the
limit $h \tau \gg 1$ {\em sign-changing oscillations} of $M(h)$
exist (the quasiperiod is still $h^* \approx \pi \eA$), although
their magnitude is suppressed in $t$ as $\sim t^2$ and
exponentially in $d/l$. On the contrary, in the opposite case $ h
\tau \ll 1$ we get that $M(h)$ {\em does not exhibit
oscillations}, being negative ($M(h)<0$) for all $h \ll 1/\tau$.
The condition $h \tau \ll 1$ (together with $l \ll d$) corresponds
to the ``diffusive'' limit of the Usadel equation and the results
obtained in this case from the Eilenberger equation can be
recovered from the Usadel equation.

We mention that nonoscillatory result for $M(h)$ in the diffusive limit
is in contrast with the behavior of the Josephson critical current $I_c(h)$
in SFS junctions.
Oscillations of $I_c(h)$ are not destroyed by disorder and persist
(although exponentially suppressed
 in $h/\ETh$) even in
the ``diffusive'' limit $h\tau \ll 1$. The period of these
oscillations is $h^* \sim \epsilon^*=\ETh$. These oscillations
were observed experimentally in
Refs.~\onlinecite{JosephEx1,JosephEx2,JosephEx3,JosephEx4,JosephEx5},
for further references see review articles
Refs.~\onlinecite{Golubov,Buzdin}. Thus, oscillations  of the
induced magnetization $M(h)$ turn out to be more sensitive to
disorder than those of the Josephson critical current $I_c(h)$.

Our analysis shows that for moderate bulk disorder ($l \gtrsim d$,
$l$ may also qualitatively include surface disorder of the
interface) and not too small interface transparency ($t\sim 1$)
the magnitude of oscillations is still quite noticeable, thus
giving hope for experimental check of our predictions. Since for
$h \tau \gg 1 $ oscillations of $M(h)$ are sustained, oscillatory
behavior of $M(h)$ should be attainable even in the presence of
disorder in the case of sufficiently strong ferromagnets. Since
the exchange field $h$ is hardly variable in the experiment, one
may hope to observe the oscillations of $M(h) $ performing
measurements on samples with different thickness $d$. We also note
that the case of thin ferromagnetic films $d \ll \xi_S$ is the
most interesting for experiment: experimentally  relevant exchange
fields are $h \gg T_c$, one needs $h \sim \eA$ to observe
oscillations, thus $d/\xi_S \sim T_c/\eA \sim T_c/ h \ll 1$.

As a limiting case of our analysis we obtain
that in the clean case for small exchange fields ($h \ll t \eA$)
and a thin ferromagnetic film ($d \ll \xi_S$)
the induced magnetization $M(h)$ is given by the universal result Eq.~(\ref{eq:Msmallh}).
This complements the analysis of Refs.~\onlinecite{BVE}
and suggests that this result holds in perturbative regime (\ref{eq:pert})
in SF systems with arbitrary strength of potential disorder.

The paper is organized as follows. In Sec.~\ref{sec:SFspectrum} we
provide qualitative quasiclassical description of Andreev spectrum
in SF systems and show how the conditions
(\ref{eq:pert}),(\ref{eq:nonpert}) and the periodicity of the
spectrum arise. In Secs.~\ref{sec:clean} we consider the limit of
clean sample and perfect interface transparency, when the
oscillatory behavior of induced magnetization is most pronounced.
We present the system and formalism of the Eilenberger equation
used to derive the expression for the induced magnetization and
analyse this expression in detail. The connection between the
predicted effect and the properties of Andreev spectrum is
discussed. In Sec.~\ref{sec:lowTrcy}  we show that our assumptions
about ideal transport properties (perfect interface transparency,
ballistic electron motion) of the system are not crucial for the
existence of oscillations. We present the results for the case of
low SF interface transparency and disordered ferromagnet and show
that oscillations exist in such limit provided some conditions on
parameters are met. Finally, we conclude with
Sec.~\ref{sec:conclusion}. In Appendix the general formulas for
the case of arbitrary interface transparency and clean samples are
given.

\section{Qualitative physics of Andreev spectrum in SF systems\label{sec:SFspectrum}}

\begin{figure}
\includegraphics
[width=0.25\textwidth]
{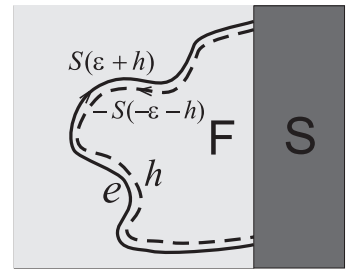}
\caption{Semiclassical description of Andreev spectrum of an SF system.
Each classical trajectory that starts and terminates at SF interface
corresponds to a set of discrete Andreev energy levels, which are obtained
from the Bohr-Sommerfeld rule Eq.~(\ref{eq:AndrSF}).}
\label{fig:SF}
\end{figure}

The spectrum of a superconductor-ferromagnet(SF) (or superconductor-normal metal(SN))
system is given by Andreev states,
which are the states of electron-hole pairs localized in  the F(N) region
due to Andreev reflection at SF(SN) interface.
A qualitative understanding of the properties of Andreev spectrum in SF systems
can be obtained from the following semiclassical picture (Fig.~\ref{fig:SF}).
We assume the case of ideal SF interface transparency here for simplicity.
(Qualitative analysis of the present section generalizes
the discussion of Andreev spectrum of SN systems done in Sec. II of Ref.~\onlinecite{TSandAA}
to SF systems.)

First, consider an SN system without exchange field.
Suppose an electron with an energy $\epsilon$ relative to the Fermi level $\epsilon_F$
inside the superconducting energy gap $\De$ ($|\epsilon| < \De$) is
travelling along some classical path in N region.
If the electron hits SN interface,
it is reflected as a hole with the energy $-\epsilon$.
A peculiar property of Andreev reflection is that
the momentum of reflected hole is opposite
(apart from a small angular mismatch $\sim \epsilon/\epsilon_F$)
to that of the incident electron.
Therefore the reflected hole will travel along the same path as
the incident electron but in the opposite direction.
If this path hits the interface again, the hole is reflected
back as the electron.
Within the quasiclassical Bohr-Sommerfeld(BS) approach
if the total action of such process is an integer multiple of $2 \pi \hbar$,
then such electron-hole pairs forms a bound state.

Thus, each classical path that starts and terminates at SN interface
corresponds to a set of discrete Andreev levels,
energies of which can be obtained from the BS rule.
For each such path $\ga$ of the length $L_\ga$
the action of the electron traversing this path is
$
    S(\epsilon)= p(\epsilon) L_\ga,
$
where
$
    p(\epsilon)=\sqrt{ 2m (\epsilon_F + \epsilon)} \approx p_F +\epsilon/v_F
$
is the absolute value of electron's momentum.
The action takes the form
\[
    S(\epsilon)=p_F L_\ga+ \epsilon \tau_\ga
\]
where $\tau_\ga =L_\ga/v_F$ is the time of traversal of path $\ga$.
The action of the hole (a missing electron) is $-S(-\epsilon)$
and the contributions from Fermi length scales cancel each other.
The BS rule gives
\[
    S(\epsilon)-S(-\epsilon)-2\phi(\epsilon)=
    2\epsilon \tau_\ga-2\phi(\epsilon)=2 \pi n,
\]
where $ \phi(\epsilon)= \arccos(\epsilon /\De) $ is the phase of Andreev reflection
and $n$ is integer, and thus discrete Andreev levels $\epsilon=\epsilon_\ga(n)$
corresponding to path $\ga$ are determined from the equation:
\beq
    \epsilon \tau_\ga=\pi n + \phi(\epsilon).
\label{eq:AndrSN}
\eeq

Generalization to SF systems is straightforward (Fig.~\ref{fig:SF}).
In SF system one should distinguish between Andreev states with electron spins
directed along ($\ua$) and opposite to ($\da$) the exchange field $h$.
``Up'' states $\epsilon_\ua$ acquire a shift $-h$
and ``down'' states $\epsilon_\da$ acquire a shift $+h$ in the F region.
The BS rule reads($+$ and $-$ correspond to $\ua$ and $\da$, respectively):
\[
    S(\epsilon_{\ua,\da} \pm h)-S(-(\epsilon_{\ua, \da} \pm h))-
    2\phi(\epsilon_{\ua,\da})=2\pi n
\]
Note that the argument of $\phi(\epsilon_{\ua, \da})$ does not
acquire the shift $\pm h$ due to exchange field, since exchange
field is absent in the superconductor and therefore does not
affect the process of Andreev reflection. Thus, the equation for
Andreev levels in SF system reads:
\beq
    (\epsilon_{\ua,\da} \pm h) \tau_\ga = \pi n + \phi(\epsilon_{\ua,\da})
\label{eq:AndrSF}
\eeq

The time $\tau^*$ defined in Sec. \ref{sec:intro} as a time spent by electron
in the F region is the characteristic time for $\tau_\ga$.
If $h \tau^* \ll 1$, then  Andreev levels of SF system
obtained from Eq.~(\ref{eq:AndrSF}) are only slightly
different from those of the corresponding SN system with $h=0$
obtained from Eq.~(\ref{eq:AndrSN}). This condition corresponds
to perturbative regime (\ref{eq:pert}).
If $h \tau^* \gtrsim 1$, then Eq.~(\ref{eq:AndrSF}) is significantly different
from  Eq.~(\ref{eq:AndrSN}) and this condition corresponds
to nonperturbative regime (\ref{eq:nonpert}).
Further, one sees that Eq.~(\ref{eq:AndrSF}) is invariant under the
periodic translation ($h \tau_\ga \rtarr h \tau_\ga+\pi$).
Thus the solutions of Eq.~(\ref{eq:AndrSF}) are periodic:
\[
    \epsilon_{\ua,\da}(h)=\epsilon_{\ua,\da}(h+\pi/\tau_\ga)
\]
This periodicity of Andreev spectrum should reveal itself in the behavior
of macroscopic quantities of SF system as a function of parameter $h \tau^*$.
This is indeed the case for Josephson critical current in SFS junctions
and, as we show in this paper, for the induced magnetization.

\section{Ballistic case, ideal SF interface transparency\label{sec:clean}}
\subsection{System and method \label{sec:system}}
We start our analysis with the case of ideal ballistic electron transport:
absence of bulk disorder and perfect interface transparency $t=1$.
In this case the oscillations of induced magnetization are most pronounced.

We consider an SF bilayered
system pictured in Fig. 1(b): the superconductor occupies the
half space $x>d$ and the ferromagnetic layer (F-layer) of the thickness $d$
is located in the region $0<x<d$ . We assume that there is no bulk disorder
neither in the F-layer nor in the superconductor and that the interfaces are
ideal, namely, the superconductor-ferromagnet(SF) interface $(x=d)$ is
perfectly transparent and the ferromagnet-vacuum (FV) interface $(x=0)$ is
specular.

We study the problem solving the Eilenberger equation
\cite{Eilenberger} for the quasiclassical Green's function
$\check{g}(\omega ,\bm{n},\bm{r})$. Here $\omega =\pi T(2m+1)$ is
the fermionic Matsubara frequency ($T$ is the temperature and $m$
is an integer), $\bm{n}$ is the unit vector representing the
direction of the electron momentum on the Fermi surface and
$\bm{r}$ is the radius vector. Due to the geometry considered
$\check{g}(\omega ,\bm{n},\bm{r})=\check{g}(\omega ,n_{x},x)$,
where $n_{x}$ is the projection of $\bm{n}$ on the $x$-direction,
$-1\leq n_{x}\leq 1$. The Green's function $\check{g}(\om,n_{x},x)$
is a matrix in the tensor product of Gor'kov-Nambu and
spin spaces:
\[
    \check{g}=\hat{g}_{1}\circ \tau _{1}+\hat{g}_{2}\circ \tau _{2}+\hat{g}_{3}\circ \tau _{3},
\]
\beq
    \hat{g}_{i}=g_{i}^{0}\hat{1}+g_{i}^{z}\hat{\sigma}_{z}.
\label{eq:gi}
\eeq
Here $\tau _{i}$, $i=1,2,3$ are the Pauli matrices in the
Gorkov-Nambu space, and $\hat{1}$, $\hat{\sigma}_{z}$ are the
unity and Pauli matrices in the spin space ($z$ denotes the
direction of the exchange field in the spin space). Diagonal representation
(\ref{eq:gi}) in the spin space is possible in the case of homogeneous magnetization,
which is assumed here.

The quasiclassical approach implies that the following conditions are
satisfied: $h\ll \epsilon_F$, $p_F d\gg 1$, where $\epsilon_F$, $p_F$
are the Fermi energy and momentum.
However, we stress that these conditions are required only
for the applicability of the method used,
but not for the very existence of the oscillations.
The results are still valid qualitatively
 in the case of strong ferromagnet of atomic thickness
($h\ \lesssim \epsilon_F$, $p_F d \gtrsim 1$),
although the effect is reduced.
The same concerns possible mismatch of electronic properties
(such as density of states $\nu_F$, Fermi velocity $v_F$)
in the ferromagnet and superconductor. It is assumed here that they are the same.

The density of the induced spin magnetization can be expressed in terms of the
quasiclassical Green's function in the following way:
\beqar
    M(h,x) &=& \mu_B (\lan \psi_\ua^+(x) \psi_\ua(x)\ran -
        \lan \psi_\da^+(x)\psi_\da(x) \ran =   \\
        &=&\frac{2\pi}{i} \mu_B \nu_F T\sum_\om\int_0^1 dn_x
        \,g_3^z(\om ,n_x,x),
\label{eq:M}
\eeqar
where $\mu_B$ is the Bohr magneton and $\nu_F$ is the electron density of states
at Fermi level per single spin projection.
The total induced magnetic moment of the system (per unit square in the
plane parallel to SF interface) is obtained by the integration of $M(h,x)$
with respect to $x$:
\beq
    M(h)=\int_{0}^{+\infty }dx\,M(h,x).
\label{eq:Mh}
\eeq

The relation between the intrinsic magnetic moment of the ferromagnet $M_{F0}$
and the exchange field $h$ acting on the electrons depends on the model of the ferromagnet
\cite{Aharoni}.
Usually $M_{F0}=M_{F0el}+M_{F0loc}$ is combined of the contribution $M_{F0el}= \chi_N h d$
produced by free electrons and the contribution $M_{F0loc}= \al \chi_N h d$ produced
by localized magnetic moments ($\al$ is some phenomenological constant and $\chi_N$ is
the normal metal spin susceptibility). In the case of itinerant ferromagnet
 the localized
magnetic moments are absent ($\al=0$) and $M_{F0}=M_{F0el}=\chi_N h d$.

We emphasize that magnetic moment $M(h)$ Eq.~(\ref{eq:Mh})
expressed in terms of the quasiclassical Green's function is the
magnetic moment \emph{induced by the presence of
superconductivity} in the system. It is determined by the
properties of the energy spectrum on the scale $\sim T_c$ near the
Fermi level. It does not include the part $M_{F0el}=\chi_N hd$ of
the intrinsic magnetic moment $M_{F0}$ of the ferromagnet produced
by free electrons that originates from the energy shift of the
entire  electron band. The latter cannot be taken into account
within the quasiclassical approach and should be added separately.
The total magnetic moment of the system is given by
Eq.(\ref{eq:Mtot}).

The Eilenberger equation for the system without disorder reads:
\begin{equation}
    v_F n_x  \partial_{x} \gc+\lt[
    (\om \hat{1}-ih(x)\hat{\sig}_{z})\circ \tau _{3}+
    \De(x)\hat{1}\circ \tau _{2},\gc \rt] =0,
\label{eq:Eilen}
\end{equation}
where $h(x)$ is the exchange field in energy
units, $\Delta (x)$ is the superconducting order parameter, and brackets $[,]
$ stand for the commutator. The exchange field is contained in the F-layer
only and assumed to be constant within the layer: $h(x)=h,$ if $0<x<d$,
and $h(x)=0,$ if $x>d$. We assume there is no BCS interaction between
electrons in the F-layer and therefore we always have $\Delta (x)=0$ for
$0<x<d.$

In principle, the exact order parameter $\Delta (x)$ has to be found
self-consistently from the solution of the Eilenberger equation (\ref{eq:Eilen})
supplemented by the self-consistency equation for the
superconducting order parameter. The order parameter $\Delta (x)$ approaches
a bulk BCS value $\Delta $ at large distances from the F-layer $x\gg \xi _{S}$,
but is partially suppressed near the F-layer. Computation of $\Delta (x)$
self-consistently is a hard analytical problem. Fortunately, our main result
about the oscillatory sign-changing behavior of $M(h)$ is not sensitive to
the exact shape of $\Delta (x)$. Therefore, we assume that $\Delta(x)=\Delta$
for all $x>d$ and perform calculations under this assumption.

Eq. (\ref{eq:Eilen}) must be supplied by proper boundary conditions at SF
and FV interfaces\cite{Zaitsev}. The limit of the ideal transparency of SF ($x=d$)
interface implies that the Green's function is continuous:
\[
    \gc(\om ,n_x,x=d-0)=\gc(\om,n_x,x=d+0).
\]
At FV ($x=0$) interface the specular reflection condition
reads:
\[
    \gc(\om ,n_x,x=0)=\gc(\om ,-n_x,x=0).
\]
At $x \gg \xi_S $
the solution approaches the BCS bulk result:
\[
    g_i^z=g_1^0=0, \mbox{ } g_2^0=f_S=\frac{\De}{\sqrt{\om^2+\De^2}},
        \mbox{ } g_3^0=g_S=\frac{\om}{\sqrt{\om^2+\De^2}}.
\]
\subsection{Analysis \label{sec:analysis}}

Under the made assumptions the solution of Eq.~(\ref{eq:Eilen}) is
straightforward and we obtain:
\begin{widetext}
\beq
g_3^z(\om, n_x, x) = -\frac{i}{2}  \fBCS^2
 \frac{ \sin 2 H}
 { \lt[ \cosh \Om
  +|\gBCS| \sinh \Om
  \rt]^2
 \cos^2 H +
 \lt[ \sinh \Om
  +|\gBCS|
  \cosh \Om
  \rt]^2
 \sin^2 H
 }
\lt\{ \begin{array}{l} 1 \mbox{,  } 0<x< d,\\
     \exp \lt( -\frac{\sqrt{\om^2+\De^2}}{\eA} \frac{x/d-1}{|n_x|} \rt)
   \mbox{,  } x> d.\\
   \end{array}
  \rt.
\label{eq:g3z}
\eeq
\end{widetext}
Here $\Omega =|\omega |/(\epsilon _{A}|n_{x}|)$, $H=h/(\epsilon _{A}|n_{x}|)$
and $\epsilon _{A}=v_{F}/(2d)$ is the Andreev energy: $\hbar /\epsilon _{A}$
is the time the electron travels from SF interface and back within F-layer
with the velocity perpendicular to the interface.

Inserting Eq.~(\ref{eq:g3z}) into Eq.~(\ref{eq:M}) and integrating over $x$
one obtains $M(h)$. The key point of our analysis is that
$ig_{3}^{z}(\omega,n_{x},x)$ is a periodic sign-changing (note $\sin 2H$ in the numerator in
Eq.~(\ref{eq:g3z})) function of $H=h/(\epsilon _{A}|n_{x}|)$ and depends on
the exchange field solely via this parameter.

General properties of $M(h)$ can be summarized as follows:
1) $M(h)$ depends on the strength of exchange field solely via the combination
$h/\epsilon_{A}=2hd/v_{F}$;
2) for any temperature $T<T_{c}$ and any ratio $d/\xi _{S}$
the induced magnetic moment $M(h)$ is an oscillating sign-changing function
of $h/\epsilon _{A}$ with a quasiperiod $h^{\ast }/\epsilon _{A}\approx \pi $,
the amplitude of the oscillations decays monotonically as $h/\epsilon _{A}$ increases;
3) hence, $M(h)$ can either have the same ($M(h)>0$)
or opposite ($M(h)<0$) direction as $M_{F0}$, depending on $h/\epsilon _{A}$;
4) at $h/\epsilon _{A}\ll 1$ we get $M(h)<0$, which indicates the
screening of $M_{F0}$; 5) the magnitude of oscillations of $M(h)$ is
largest at $T=0$ and decreases  as $T$ increases;
$M(h)=0$ at $T=T_c$.

The spatial dependence of $M(h,x)$ shown in Fig. 1(b) is governed by $g_3^z(\om,n_x,x)$:
$M(h,x)$ is constant within the F-layer and decays exponentially
over the distance $\xi_S$ into the superconductor.
If $d\sim \xi _{S}$, then parts of $M(h)$ located in the F-layer
\[ M_{F}(h)=\int_{0}^{d}dx\,M(h,x) \]
and in the superconductor \[ M_{S}(h)=\int_{d}^{+\infty }dx\,M(h,x) \]
are of the same order. If $d\gg \xi _{S}$, then the induced magnetic moment
is located predominantly in F-layer and
$M(h)\approx M_{F}$, $M_{S}/M_{F}\sim \xi _{S}/d\ll 1$.
In the opposite limit $d\ll \xi _{S}$ the induced magnetization is located mainly in the
region of the superconductor of the size $\xi _{S}$ near the F-layer
and $M(h)\approx M_{S}$, $M_{F}/M_{S}\sim d/\xi _{S}\ll 1$.

Below we concentrate on the experimentally more relevant situation of a thin
F-layer $d\ll \xi _{S}$ and illustrate the announced properties of $M(h)$
explicitly for this particular case. In this regime, the expression for
$M(h)\approx M_{S}(h)$ can be reduced to the form:
\begin{eqnarray}
    M(h) &=&-d\mu _{B}\nu _{F}\epsilon _{A}\pi T\sum_{\omega }
    \frac{\Delta ^{2}}{\sqrt{\omega ^{2}+\Delta ^{2}}}\times   \notag \\
    &&\times \int_{0}^{1}dn_{x}\,n_{x}\frac{\sin 2H}{\omega ^{2}+\Delta ^{2}\cos^{2}H}.
\label{eq:Mthin}
\end{eqnarray}

First, we consider the zero temperature limit $T=0$. Replacing the sum over
$\omega $ by the integral $T\sum_{\omega }\ldots =\int_{-\infty }^{\infty}
d\omega /(2\pi )\ldots $, we obtain:
\beq
    M(h)=-2d\mu _{B}\nu _{F}\epsilon _{A}\int_{1}^{\infty }\frac{dt}{t^{3}}
\,
    Z_{\pi }\left( \frac{h}{\epsilon _{A}}t\right)
\label{eq:MT0}
\eeq
where $Z_{\pi }(x)=x$, if $-\pi /2<x<\pi /2$, and periodically continued to
all $x$ (linear ``zig-zag''-type function with a period $\pi $). The
integral with respect to $n_x$ can easily be calculated and we obtain the function
shown in Fig. 1(a).

Close to superconducting transition point ($(T_c-T)/T_c\ll 1$) we obtain
\[
M(h)=-d\mu _{B}\nu _{F}\epsilon _{A}\pi T\sum_{\omega }\frac{\Delta ^{2}}
{|\omega |^{3}}\int_{1}^{+\infty }\frac{dt}{t^{3}}\sin \left(
2\frac{h}{\epsilon _{A}}t\right).
\]
We see
that $M(h)$ is again an oscillating sign-changing
function of $h/\epsilon _{A}$ with quasi-period $h^{\ast }/\epsilon
_{A}\approx \pi $, although the amplitude of oscillations is parametrically
smaller than that at  $T=0$ by
$(\Delta(T)/T_{c})^{2}\sim (T_{c}-T)/T_{c}\ll 1$.

\subsection{Perturbative regime ($ h \ll t \eA$) \label{sec:smallh}}
In the limit $h\ll \epsilon_A$ from Eq.~(\ref{eq:Mthin}) we get:
\beq
    M(h) =-\chi_N \pi T\sum_{\omega }
    \frac{\De^2}{(\om^2+\De^2)^{3/2}} h d
\label{eq:Msmallh2}
\eeq
where $\chi_N=2 \mu_B \nu_F$ is the bulk spin susceptibility of the normal metal.
Since
\beq
    \chi_S(T)=
        \chi_N \lt(1-\pi T\sum_{\omega }
            \frac{\De^2}{(\om^2+\De^2)^{3/2}} \rt)
\label{eq:chiS}
\eeq
is the bulk spin susceptibility of the superconductor,
Eq.~(\ref{eq:Msmallh2}) can be rewritten in the form Eq.~(\ref{eq:Msmallh}).
From the formulas given in Appendix one obtains the same result in the case
of clean samples and low SF interface transparency ($t \ll 1$), provided
that $h \ll t \eA $.
The total magnetic moment produced by free electrons is
$M_{F0el}+M(h)= \chi_S(T) hd $. At zero temperature $\chi_S(T=0)=0$
and $M(h)=-M_{F0el}= \chi_N hd $, i.e. the induced
magnetic moment $M(h)$ totally screens the part $M_{F0el}$
of the intrinsic moment $M_{F0}$ produced by free electrons.
And interesting feature is that $M_{F0el}$ is located in the ferromagnet,
whereas $M(h)$ is spread over the distance $\xi_S$ from the F-layer in the
superconductor. Since for $h\ll \epsilon_A$ and $d\ll \xi_S$ the exact
order parameter $\Delta (x)$ is only slightly suppressed due to the
ferromagnetic proximity effect, this result is justified even if the
self-consistency condition for $\Delta (x)$ is taken into account.

As the same result Eq.~(\ref{eq:Msmallh}) was obtained in the opposite diffusive
limit
 for $h \ll \ETh $, we make a conjecture that Eq.~(\ref{eq:Msmallh})
holds for arbitrary strength of potential disorder, provided the general
condition Eq.~(\ref{eq:pert}) is met. The universality of result Eq.~(\ref{eq:Msmallh})
is reminiscent of the properties of the bulk linear spin susceptibility
of the superconductor $\chi_S(T)$ (Eq.~(\ref{eq:chiS})).
It is also independent of
the strength of potential disorder \cite{AG}.

\subsection{Andreev states \label{sec:Andreev}}
The oscillations of induced magnetization are closely related to the
properties of the energy spectrum of localized Andreev states in the system \cite{AndreevSF}.
The equation for Andreev energy levels $\epsilon_{\ua ,\da }$
with the electron's spin having the same $(\ua)$ and opposite $(\da)$
direction as the exchange field (corresponding to $+$ and $-$ signs, respectively) reads:
\beq
       \frac{\epsilon_{\ua,\da}\pm h}{\epsilon_A |n_x|}
        -\arccos \frac{\epsilon_{\ua,\da}}{\De } =\pi n,
\label{eq:AndrSFclean}
\eeq
where $n$ is integer.
Note that in the case of ballistic system this equation
exactly coincides with Eq.~(\ref{eq:AndrSF}) used for qualitative considerations.
Eq.~(\ref{eq:AndrSFclean}) is invariant under the periodic translation
$h/(\eA|n_x|)\rtarr h/(\eA|n_x|)+\pi k$ ($k$ is integer)
in the same fashion as the Green's function (\ref{eq:g3z}) is periodic in $H$.
In the limit $d\ll \xi _{S}$ there exists only one level for a given $n_{x}$ and
projection of spin
\footnote{This statement and the expression Eq.~(\ref{eq:Andr}) are actually violated
when $H$ is close to $\pi n$, however this is not important for present
considerations.}:
\begin{equation}
    \epsilon_{\ua,\da }(H)=\pm \De \cos H,\mbox{ }H\in [ 0,\pi ],  \label{eq:Andr}
\end{equation}
and periodically continued to all $H$. The states with
$H\in [0,\pi/2]+\pi n$ ($\epsilon_\ua>0$, $\epsilon_\da<0$)
contribute to the screening of $M_{F0}$
($\sin 2H>0$, see Eq.~(\ref{eq:Mthin})),
whereas the states with
$H \in [\pi /2,\pi ]+\pi n$ ($\epsilon_\ua<0$, $\epsilon_\da>0$)
give rise to the anti-screening of $M_{F0}$ ($\sin 2H<0$). Due to the property
\[
\epsilon_{\ua,\da}(\pi /2+\pi n+\delta H)=
\epsilon_{\da,\ua}(\pi /2+\pi n-\delta H),
\]
$\delta H\in [0,\pi /2]$,
such ``up'' and ``down'' states interchange
in the energy space but since the spin direction is ``attached'' to them
explicitly, this results in the opposite signs of contributions to $M(h)$.

\subsection{Self-consistency of order parameter $\De(x)$ \label{sec:Delta}}
As it has been mentioned, the oscillating behavior of $M(h)$ is insensitive
to the exact shape of $\Delta (x)$ and therefore persists if the
self-consistency of $\Delta (x)$ is taken into account.
This is the case, because the periodic functions of $H$ in
Eq.~(\ref{eq:g3z}) arise from the solution of Eq.~(\ref{eq:Eilen})
in the F-layer, where $\Delta (x)=0$ and the general solution can
always be found explicitly. Due to this fact $ig_{z}^{3}$ is a
periodic sign-changing function of $H$ and, hence, $M(h)$ is a
quasiperiodic sign-changing function of $h/\epsilon _{A}$,
independently of the exact shape of $\Delta (x)$. Since for
arbitrary $T$ the proof is cumbersome, we illustrate it here in
the simplest case, when $T$ is close to $T_{c}$. In this limit one
can obtain the solution to Eq.~(\ref{eq:Eilen}) as an expansion in
$\Delta (x)$. In the lowest (quadratic) order we find for $x>d$:
\beqar
    g_{3}^{z}(\omega ,n_{x},x) &=&-\frac{i}{2\xi ^{2}}\sin \left( 2\frac{h}{\epsilon _{A}}
    \frac{1}{|n_{x}|}\right) \frac{1}{\omega ^{2}}\times  \\
    &\times &\int_{x}^{\infty }\Delta (x^{\prime })e^{-\frac{x^{\prime }}{\xi }}
    \,dx^{\prime }\int_{d}^{\infty }\Delta (y)e^{-\frac{y}{\xi }}\,dy ,
\eeqar
where $\xi =\frac{v_{F}|n_{x}|}{2|\omega |}$. This yields the form
\[
M(h)=-\int_{1}^{\infty }dt\,\sin \left( 2\frac{h}{\epsilon _{A}}t\right) F(t,h),
\]
where $F(t,h)$ is a \emph{positive monotonically decreasing} with respect to
$t$ envelope function. The above integral can be both positive and negative
depending on the value of $h/\epsilon _{A}$. This is especially clear for
$h/\epsilon _{A}\gg 1$ when one can integrate by parts to obtain
\[
     M(h)\approx \frac{\cos \left( 2h/\eA\right)}{ 2h/\eA}F(1,h).
\]
Therefore the induced magnetization is an oscillating function of
the parameter $ 2h/\eA\ $ regardless of the
exact form of $\De(x)$.

\section{Low SF interface transparency \label{sec:lowTrcy}}

Our assumptions about ideal transport properties of the system
(perfect SF interface transparency, ballistic electron motion in
the samples) are not crucial for the existence of oscillations of
induced magnetization. The oscillations of $M(h)$ exist for
arbitrarily low SF-interface transparency (see Appendix).
Moreover, they can exist in the presence of bulk disorder in the
sample.

To illustrate that we turn to the case of low SF interface transparency $t=t(n_x) \ll 1$
($t(n_x)$ is a transmittance coefficient, see Appendix).
In this limit the proximity effect is weak and
one can take the effect of disorder into account
by linearizing the Eilenberger equation with collision term \cite{BVE01,Fominov}.
We assumed that superconductor is clean and ferromagnet is disordered
and described by the mean free path $l$ and scattering time $\tau=l/v_F$.

The Eilenberger equation in the F region ($0<x<d$)
for the ``up'' component $\gc_\ua=\gc$ reads
(we omit the index $\ua$ here for brevity):
\beq
    v_F n_x \pd_x \gc+\lt[ (\om- i h ) \tau_3+
        \frac{1}{2 \tau} \lan \gc \ran, \gc \rt] = 0
\label{eq:EilenF}
\eeq
where $\lan \gc \ran=1/2 \int_{-1}^{1} d n_x \gc(n_x ,x) $
is the angular averaging.
In the S region ($x>d$):
\beq
   v_F n_x  \pd_x \gc+\lt[
    \om \tau_3+\De \tau_2,\gc \rt] =0.
\label{eq:EilenS}
\eeq
In the zeroth order in interface transparency ($t(n_x)=0$)
the superconductor and ferromagnet are not linked and the solution is:
\[
    \gc^{(0)}_F= \sgn \om \tau_3 \mbox{, } 0<x<d,
\]
\[
    \gc^{(0)}_S= f_S \tau_2 + g_S \tau_3 \mbox{, } x>d
\]
Next, we present the Green's function in the ferromagnet in the form
\[
    \gc=\gc^{(0)}_F+\de \gc, \mbox{ } \de \gc = \de g_1 \tau_1 + \de g_2 \tau_2 +\de g_3 \tau_3,
\]
and, leaving only linear in $\de \gc$  terms in Eq.~(\ref{eq:EilenF}),
arrive at the following equation for $\de \gc$:
\beq
    v_F n_x \pd_x \de \gc+
    \lt[ \lt( \om -i h +\frac{ \sgn \om }{2 \tau} \rt) \tau_3, \de \gc \rt]+
    \frac{\sgn \om}{2\tau} [\lan \de \gc \ran, \tau_3 ]= 0.
\label{eq:EilenFdg}
\eeq
First we need to obtain a linear in $t$ solution for $\de g_2$ in the F region.
From Eq.~(\ref{eq:EilenF}) we get:
\beq
    l^2 n_x^2 \pd_x^2 \de g_2 - \al_\om^2 \de g_2=- \al_\om \lan \de g_2 \ran
\label{eq:dg2}
\eeq
where $\al_\om = 1+2(|\om|-i h \sgn \om) \tau$.
Boundary condition with vacuum reads:
\beq
    \pd_x \de g_2 (x=0)=0
\label{eq:dg2FV}
\eeq
At the SF interface one must use Zaitsev
boundary conditions (\ref{eq:ZaitsevBC}) for nonideal interface
transparency. In the limit $t \ll 1$ one can expand them in $t$.
First order in $t$ gives:
\[
    \de g_1(n_x, x=d-0)= i\frac{t}{2} \sgn \om \sgn n_x f_S.
\]
Using the $\tau_2$-component of Eq.~(\ref{eq:EilenFdg})
\[
    l n_x \pd_x \de g_2 + i \al_\om \sgn \om \de g_1 = 0,
\]
we arrive at the boundary condition for $\de g_2$ at SF interface:
\beq
    \pd_x \de g_2 (x=d-0) = \frac{t(n_x)}{2 l |n_x|} f_S \al_\om.
\label{eq:dg2SF}
\eeq
Eq.~(\ref{eq:dg2}) must be solved for $x \in[0,d]$ with boundary conditions
Eq.~(\ref{eq:dg2FV}),(\ref{eq:dg2SF}).
Condition Eq.~(\ref{eq:dg2FV}) allows to symmetrically continue $\de g_2$ to
$[-d,0]$ interval.
We perform the Fourier transformation\cite{BVE01,Fominov}:
\[
    \de g_2 (x)= \sum_{n=-\infty}^{+\infty} \de g_2 (n) e^{i k_n x},
\]
where $ k_n = \pi n /d$ ($n$ is integer) and Fourier coefficients are
\[
    \de g_2 (n)= \frac{1}{2d} \int_{-d}^{d} \de g_2(x) e^{-i k_n x} dx.
\]
Calculating Fourier coefficient of both sides of
Eq.~(\ref{eq:dg2}) and taking the boundary condition
(\ref{eq:dg2SF}) into account, we get:
\beq
    \de g_2 (n)= \frac{\al_\om}{L_n}
        \lt( \lan \de g_2(n) \ran+(-1)^n \eA \tau |n_x| t(n_x) f_S \rt),
\label{eq:dg2n}
\eeq
where $L_n=l^2 n_x^2 k_n^2 + \al_\om^2$.
Angular averaging of Eq.~(\ref{eq:dg2n}) gives:
\[
    \lan \de g_2(n) \ran =
        \frac{(-1)^n \al_\om f_S \eA \tau  \lan \frac{|n_x| t}{L_n} \ran}
        {1-\al_\om \lan \frac{1}{L_n} \ran }
\]
and thus
\[
    \de g_2(n) =  (-1)^n \de g_2^*(n),
\]
\beq
    \de g_2^*(n) = \frac{\al_\om}{L_n} \eA \tau  f_S
        \lt( \frac{ \al_\om \lan \frac{|n_x| t}{L_n} \ran}
        {1-\al_\om \lan \frac{1}{L_n} \ran } +|n_x| t  \rt).
\label{eq:dg2n}
\eeq

Next we solve Eq.~(\ref{eq:EilenS}) in the S region and get for $\de \gc= \gc - \gc_S^{(0)}$:
\beq
    \de \gc = c(n_x) ( \sgn n_x \tau_1 - g_S \tau_2 +f_S \tau_3)
        e^{-\frac{2 \sqrt{\om^2+\De^2} (x-d)}{v_F |n_x|} }
\label{eq:dgS}
\eeq
where $c(n_x)$ is a symmetric (yet unknown) function of $n_x$.

It follows from Eqs.~(\ref{eq:M}),(\ref{eq:dgS}) that the induced magnetic moment is given by
(again we assume the case of a thin F film ($d \ll \xi_S$) and therefore
the induced magnetic moment $M(h) \approx M_S(h)$ is located in the superconductor):
\[
    M(h)= \mu_B \nu_F 2 \pi T \sum_\om \int_0^1 d n_x
        \frac{v_F n_x}{\sqrt{\De^2+\om^2}} f_S \Imp c(n_x)
\]
Expanding Zaitsev's boundary conditions to the second order in $t$, we obtain:
\[
   \Imp c(n_x)= - \frac{t(n_x)}{2} g_S \Imp \de g_2(n_x, x=d-0)
\]
and therefore
\beqar
    M(h) & = &  - \mu_B \nu_F v_F \pi T \sum_\om \frac{  f_S g_S  }{\sqrt{\De^2+\om^2}} \times\\
    & \times & \int_0^1 d n_x  n_x t(n_x) \Imp  \de g_2(x=d).
\eeqar
The needed quantity is
\[
    \de g_2 (x=d)= \sum_{n=-\infty}^{\infty} (-1)^n \de g_2 (n)
        = \sum_{n=-\infty}^{\infty} \de g^*_2 (n).
\]

Below we analyse two different limiting cases depending on the strength of disorder.
It appears that the influence of disorder on
behavior of induced magnetization $M(h)$ is governed by
parameter $h\tau$ rather than $d/l$.

\subsection{Quasiballistic case ($h \tau \gg 1$)}
If $h \tau \gg 1$, then $\al_\om / L_n \sim 1/\al_\om \sim 1/(h \tau) \ll 1$
and one can neglect the first term in parentheses in Eq.~(\ref{eq:dg2n}) and get:
\[
    \de g_2^*(n) = \frac{\al_\om}{L_n} \eA \tau  f_S  |n_x| t
\]
Summing the series, we get
\[
    \de g_2(n_x, n)=\frac{t}{2} f_S
    \coth \lt( \frac{|\om|-ih \sgn \om+ \frac{1}{2\tau}}{\eA |n_x|} \rt)
\]
and
\beqar
    M(h) &=& - \frac{1}{2} \mu_B \nu_F v_F
    \pi T \sum_\om \frac{  f^2_S g_S  }{\sqrt{\De^2+\om^2}} \times \\
    &\times &   \int_0^1 d n_x  n_x t^2(n_x)
    \Imp \coth \lt( \frac{|\om|-ih \sgn \om+\frac{1}{2\tau}}{\eA |n_x|} \rt).
\eeqar

For not too small disorder ($d/l \gtrsim 1$)
(and a thin F-layer ($d \ll \xi_S$)
we get:
\beqar
    M(h) & = & - \frac{1}{2} \mu_{B} \nu_{F} v_F
        \pi T \sum_\om \frac{\De^2 |\om|}{(\om^2+\De^2)^2} \times \\
      & \times &  \int_0^1 d n_x n_x t^2(n_x)
    \exp \lt(-\frac{2d}{n_x l} \rt) \sin 2H.
\eeqar
One sees that oscillatory behavior is sustained, although
the magnitude of oscillations is suppressed in $t(n_x) \ll 1$ and $l/d$.
Extrapolation of this formula to not too small $t \sim 1$ gives
that for moderate disorder $ l \gtrsim d $ the magnitude
of oscillations is quite comparable to the ideal case Eq.~(\ref{eq:Mthin}).
We also mention here, that spin-orbit scattering should also suppress oscillations
of $M(h)$. Spin-orbit scattering is neglegible, if the corresponding mean free path $ l_{so} \gg d$.
This condition is always satisfied for $l \gtrsim d$, because $ l_{so} \gg l$.

\subsection{Diffusive case ($ h \tau \ll 1 $)}

If $h \tau \ll 1$ and $l \ll d$, then $\al_\om \lan 1/L_n \ran
\rtarr 1$ and the main contribution to  $\de g_2^*(n)$ comes from
the first term in parentheses in Eq.~(\ref{eq:dg2n}), which has a
(diffusion) pole, and the second term can be neglected:
\[
    \de g_2^*(n) =  \eA  f_S \lan  |n_x| t \ran \frac{1}{2(|\om|-i h \sgn \om)+D k_n^2}
\]
where $D=v_F l/3$ is the diffusion coefficient.
Summing the series, we get
\[
    \de g_2(x=d)=
       f_S \lan  |n_x| t \ran \frac{3d}{2l}
        \frac{\coth \sqrt{\frac{2(|\om|-ih \sgn \om)}{\ETh}}}
            {\sqrt{\frac{2(|\om|-ih \sgn \om)}{\ETh}}},
\]
and
\beqar
    M(h) &=& - \mu_B \nu_F v_F \pi T \sum_\om \frac{  f^2_S g_S  }{\sqrt{\De^2+\om^2}}
    \times \\ & \times &    \lan |n_x| t(n_x)\ran^2  \frac{3d}{2l}
        \Imp \frac{\coth \sqrt{\frac{2(|\om|-ih \sgn \om)}{\ETh}}}
            {\sqrt{\frac{2(|\om|-ih\sgn \om)}{\ETh}}},
\eeqar
where $\ETh=D/d^2$ is the Thouless energy.
This result is valid, if $t \ll l/d $ and  $h \gg t \eA$.
Since the conditions $h \tau \ll 1$ and  $l \ll d$
correspond to ``diffusive'' regime, this result could be obtained from the Usadel equation.

Interestingly, $\Imp \de g_2(x=d)$ {\em does not oscillate}
as a function of $h$, even though it contains trigonometric functions.
Therefore, we obtain that in the diffusive limit $h \tau \ll 1$
the induced magnetization $M(h)$ is not oscillatory and always negative.

\section{Conclusion \label{sec:conclusion}}
In conclusion, we have shown that in SF systems the total spin magnetic moment in
the superconducting state can be both smaller and larger than that in the normal
state. The effect is due to peculiar periodic properties of Andreev states in
SF systems that result in  oscillatory sign-changing behavior of the
superconductivity-induced magnetization of the system. The predicted effect
is expected to be best observable in relatively clean SF systems
with good quality of interfaces.
Practically this means that the mean free path $l$ should be larger
than the ``exchange length'' $l_{\text{exc}} = v_F/h$. This condition
can be fulfilled in the case of sufficiently strong ferromagnets. On the other hand
$l$ should not be much smaller than the thickness of the ferromagnetic
film $d$.
We ignored a change in the magnetic moment $M(h)$ caused
by the Meissner currents assuming that the thicknesses
of the ferromagnet and superconductor are
smaller than the London penetration length.
In this case the contribution of these currents to $M(h)$ is small.
Spontaneous orbital effects in clean SF structures were studied in Ref.~\onlinecite{Annett}.

We would like to thank \textit{SFB 491 Magnetische Heterostrukturen}
and \textit{German-Israeli Foundation for Scientific Research and Development
}
for financial support.

\appendix
\section{Clean samples, arbitrary SF interface transparency}

Boundary conditions for nonideal interface transparency
have been derived by Zaitsev\cite{Zaitsev}.
They are expressed in terms of the antisymmetric
\[
    \ac(n_x,x)=(\gc(n_x,x)-\gc(-n_x,x))/2
\]
and symmetric
\[
    \sch(n_x,x)=(\gc(n_x,x)+\gc(-n_x,x))/2
\]
parts of Green's function in the following way:
\beq
    \ac((1-t)(\sch_+ +\sch_-)^2+(\sch_+ -\sch_-)^2)=t(\sch_+-\sch_-)(\sch_+ +\sch_-).
\label{eq:ZaitsevBC}
\eeq
Here $\ac=\ac(n_x,d)$, $\sch_+=\sch(n_x,d+0)$, $\sch_-=\sch(n_x,d-0)$.
Antisymmetric part $\ac(n_x,d)$ is continuous
 at the boundary $x=d$.
The transmittance coefficient $t(n_x)$ can vary from $t(n_x)=0$ for nontransparent
interface (e.g. boundary  with vacuum) to $t(n_x)=1$ for perfectly transparent interface.

In the case of clean samples and arbitrary transparency $t(n_x)$ one must
solve Eilenberger equation (\ref{eq:Eilen}) with boundary conditions (\ref{eq:ZaitsevBC}).
We obtain ($\om>0$):
\begin{widetext}
\beq
g_3^z(\om, n_x, x) = i \fBCS^2 \text{Im}
 \frac{\sinh(\Om-i H)}
 {\sqrt{
    \lt[ \gBCS \cosh(\Om-i H)+
        \lt( 2/t(n_x)-1 \rt) \sinh(\Om-i H)
    \rt]^2
     +\fBCS^2
 }}
     \exp \lt( -\frac{\sqrt{\om^2+\De^2}}{\eA} \frac{x/d-1}{|n_x|} \rt)
\eeq
in the superconductor ($x>d$) and
\end{widetext}
\begin{widetext}
\beq
g_3^z(\om, n_x, x) = 2 i \text{Im}
 \frac{\gBCS \cosh(\Om-i H)+
        \lt( 2/t(n_x)-1 \rt) \sinh(\Om-i H) }
 {\sqrt{
    \lt[ \gBCS \cosh(\Om-i H)+
        \lt( 2/t(n_x)-1 \rt) \sinh(\Om-i H)
    \rt]^2
     +\fBCS^2}}
\eeq
\end{widetext}
in the ferromagnet ($0<x<d$) (For notation, see Secs.~\ref{sec:system},\ref{sec:analysis}).
One can check that induced magnetization $M(h)$ following from these formulas
is an oscillating sign-changing function for arbitrary $t(n_x)$, $0<t(n_x)\le 1$.
The magnitude of oscillations of $M(h)$ is greatest at $t(n_x)=1$ and
decreases as $t(n_x)$ decreases; $M(h)=0$ at $t(n_x)=0$.

\end{document}